\documentclass[prl,aps,floats,floatfix,superscriptaddress,twocolumn,showpacs,letterpaper]{revtex4}

\usepackage[dvips]{graphicx}
\usepackage{amsmath} 
\usepackage{psfrag}
\usepackage{longtable}
\usepackage{dcolumn,epsfig}

\def\laq{\raise 0.4ex\hbox{$<$}\kern -0.8em\lower 0.62ex\hbox{$\sim$}}
\def\gaq{\raise 0.4ex\hbox{$>$}\kern -0.7em\lower 0.62ex\hbox{$\sim$}}

\newcommand{\beq}{\begin{equation}}
\newcommand{\eeq}{\end{equation}}
\newcommand{\bea}{\begin{eqnarray}} 
\newcommand{\eea}{\end{eqnarray}}
\newcommand{\ba}{\begin{array}}
\newcommand{\ea}{\end{array}}

\newcommand{\comment}[1]{}

\newlength{\sizeonefig}
\newlength{\sizetwofig}
\newlength{\sizeonefigb}
\newlength{\sizetwofigb}
\begin{document}
\title{Displacement- and Timing-Noise Free Gravitational-Wave Detection}
\author{Yanbei Chen}
\affiliation{Max-Planck-Institut f\"ur Gravitationsphysik, Am M\"uhlenberg 1, 14476 Potsdam, Germany}
\author{Seiji Kawamura}
\affiliation{TAMA Project, National Astronomical Observatory of Japan, Mitaka, Tokyo,
Japan}
\date{February  19, 2005}
\begin{abstract}
Motivated by a recently-invented scheme of displacement-noise-free
gravitational-wave detection, we demonstrate the existence of
gravitational-wave detection schemes insusceptible to both displacement 
and timing (laser) noises, and are thus realizable by shot-noise-limited laser interferometry.  This is possible due to two
reasons: first, gravitational waves and displacement disturbances
contribute to light propagation times in different manners; second, for an
$N$-detector system, the number of signal channels is of the order
$\mathcal{O}(N^2)$, while the total number of timing- and
displacement-noise channels is of the order $\mathcal{O}(N)$.   
\end{abstract}
\preprint{AEI-2005-089}

\maketitle

In a recent Letter~\cite{KC}, we have demonstrated the possibility of displacement-noise-free gravitational-wave (GW) detection, based on the insight that GWs and test-mass displacement disturbances contribute differently to light propagation times: the former builds up gradually as the light travel from one
detector to another, while the latter contributes only at instants
the light leave or reach the detectors (as can be seen both from the TT gauge~\cite{KC}, and from the proper reference frame of a fiducial test mass~\cite{Malik}). The elimination of displacement noises in GW detection can be extremely useful: it implies that, {\it in principle,} test masses no longer need to be isolated from environmental vibrations, and that quantum fluctuations of test-mass positions will no longer affect detection sensitivity. 

Most promising ways of detecting GWs  (except in very low frequencies) use laser interferometry~\cite{LIGO,LISA}, in which light propagation times are measured effectively by comparing phases of laser light traveling between test masses.  Since the scheme in Ref.~\cite{KC} does not cancel  clock noise, it will suffer from laser noise once converted into interferometry;  this can be a major obstacle toward implementation: (i) in many situations~\cite{LIGO,LISA}, laser noise would dominate
over all other noises by several orders of magnitude, unless schemes that  cancel laser noise are used (equal-arm Michelson interferometry in LIGO and 
Time-Delay Interferometry in LISA~\cite{TDI}); (ii) even if shot-noise-limited lasers are provided, any relative motions between the laser and optical elements participating in light emission and reception will re-create laser noise via doppler shift (e.g., the ``optical-bench noise'' of LISA). 

In this Letter, we study jointly the clock and displacement
noises of a system of $N$ detectors, with communications between each
pair of them, which gives us $N(N-1)$ channels of timing signals. For such a system,
we  have $N$ timing noises, and $Nd$ 
displacement noises, where $d$ is the spatial dimensionality of the detector
configuration: $d=1,2,3$ for colinear, coplanar and for generic configurations, respectively. Altogether, we
have $N(d+1)$ channels of noise.  As a consequence, if the number of detectors is large enough, with $N>d+2$, then the number of timing-signal channels is larger than the number of noise channels, and it must be possible to construct at least $N(N-1) - N(d+1) = N(N-d-2)$ timing combinations that are insusceptible to both displacement and clock noises.  [A similar redundancy of signal channels over noise channels have been employed in Radio and Optical astronomy to eliminate antenna phase noises and atmospheric disturbances, using techniques of {\it closure phase}~\cite{PC}.] 

After deriving conditions for displacement- and laser-noise-free detection, we will first show that for {\it one dimensional} configurations, all noise-free combinations automatically have vanishing sensitivities to GWs. We then present a two dimensional configuration (shown in Fig.~\ref{fig1}), for which at least one noise-free combination has non-vanishing GW sensitivity.

{\it Noise and Signal Transfer Functions.}
As in Ref.~\cite{KC}, we work in the TT
gauge, and suppose the ideal world lines 
of the detectors are $[T,\mathbf{X}_{(j)}]$, where $j=1,2,\ldots,N$
and $\mathbf{X}_{(j)}$ are constant 3-vectors. Suppose each detector $(j)$
deviates from its ideal world line by a displacement noise, resulting in
an actual world line of $[T,\mathbf{X}_{(j)}+ \mathbf{x}_{(j)}(T)]$.  We also assume each detector to carry a clock, which indicates 
\begin{equation}
t_{(j)} = T + \tau_{(j)}(T)\,,
\end{equation}
i.e., the coordinate time plus a timing noise, $\tau_{(j)}(T)$. [Up to accuracy of $\mathcal{O}(\dot{\mathbf{x}}^2)$, the coordinate time is identical to the proper time of the detector.]   For each $1 \le j,\, k \le N$, we send pulses from $(j)$
to $(k)$, and record the emission and receipt ``times'' according
to clock readings. Suppose a pulse is
emitted from  $(j)$ at
$t_{(j)}=t_1$  (according to clock $j$), follows a null geodesic, and then reaches
the  $(k)$ at time $t_{(k)}=t_2$ (according to clock $k$), we write,
\begin{equation}
\Delta_{(jk)}(t_1) \equiv t_2 -t_1 - L_{(jk)}\,,
\end{equation} 
where $L_{(jk)}=|\mathbf{X}_{(k)}-\mathbf{X}_{(j)}|$.  After simple calculation (see, e.g., Ref.~\cite{KC}), we obtain, up to
linear order in $h$, $\mathbf{x}_{(l)}$ and $\tau_{(l)}$ ($l=1,\ldots,N$), \begin{eqnarray}
\label{eq:Delta:t}
&&\Delta_{(jk)}(t_1)\nonumber \\
 &=&
\left\{\tau_{(k)}[t_1+L_{(jk)}]-\tau_{(j)}(t_1)\right\} \nonumber \\
&+& \mathbf{n}_{(jk)}\cdot
\left\{\mathbf{x}_{(k)}[t_1+L_{(jk)}]-\mathbf{x}_{(j)}(t_1)
\right\} \nonumber \\
&+&  
\left[\mathbf{n}_{(jk)}\otimes\mathbf{n}_{(jk)}\right]:
\bigg\{ \int_0^{L_{(jk)}} dt'  \nonumber \\
 &&{ \sum_{p=+,\times}}\mathbf{e}^p h_p\Big[(t_1-\mathbf{e}_z \cdot \mathbf{X}_{(j)}) 
+ (1-\mathbf{e}_z
  \cdot \mathbf{n}_{(jk)} ) t'\Big]\bigg\},\quad
\end{eqnarray}
with 
\begin{eqnarray}
\mathbf{e}^+=\frac{\mathbf{e}_x \otimes \mathbf{e}_x - \mathbf{e}_y
  \otimes \mathbf{e}_y}{2}\,,\;\;
\mathbf{e}^\times=\frac{\mathbf{e}_x \otimes \mathbf{e}_y + \mathbf{e}_y
\otimes   \mathbf{e}_x}{2}.\,
\end{eqnarray}
In Eq.~\eqref{eq:Delta:t}, arguments of $\tau_{(k),(j)}$,
$\mathbf{x}_{(k),(j)}$ are the coordinate time $T$ at the events of
receipt and emission, instead of the readings from
clocks at these events; on the other hand, the
argument of $\Delta_{(jk)}$ is always the  clock reading at the event
of pulse emission.  This distinction makes no difference in
Eq.~\eqref{eq:Delta:t} up to $\mathcal{O}(h)$, yet it makes versions of Eq.~\eqref{eq:Delta:t}  with different pairs of $(j,k)$ logically compatible to each
other.     Going to the frequency domain, we have
\begin{eqnarray}
\label{eq:Delta:f}
\tilde \Delta_{(jk)} 
 &= &\left[e^{i\Omega L_{(jk)}}\tilde \tau_{(k)} - \tilde \tau_{(j)}\right] \nonumber \\
&+& \mathbf{n}_{(jk)} \cdot 
\left[
e^{i\Omega L_{(jk)}}\tilde{\mathbf{x}}_{(k)} - \tilde{\mathbf{x}}_{(j)}\right] \nonumber \\
&+& 
\left[\mathbf{n}_{(jk)}\otimes\mathbf{n}_{(jk)}\right] 
:\Big\{ {\textstyle \sum_p} \mathbf{e}^p \tilde h_p
\nonumber\\
&&\qquad\frac
{
e^{i\Omega [ L_{(jk)} - \mathbf{e}_z \cdot \mathbf{X}_{(k)} ]}
-
e^{i\Omega [-\mathbf{e}_z \cdot \mathbf{X}_{(j)}]}
}
{i\Omega[1-\mathbf{e}_z\cdot \mathbf{n}_{(jk)}]}
\Big\}
\,,\;
\end{eqnarray}
where $\tilde\Delta_{(jk)}(\Omega)$, $\tilde \tau_{(l)}(\Omega)$ and $\tilde {x}_{(l)}(\Omega)$ are Fourier transforms of $\Delta_{(jk)}(t)$, $\tau_{(l)}(t)$ and $\mathbf{x}_{(l)}(t)$, $l=j,k$. [We have dropped their argument $\Omega$ for simplicity.]

\comment{
As a check, 
we expand Eq.~\eqref{eq:Delta:f} to leading order in $\Omega L/c$, which gives
\begin{eqnarray}
\Delta_{(jk)} 
&=& \tilde\tau_{(k)}-\tilde\tau_{(j)} \nonumber \\
&+& \mathbf{n}_{(jk)} \cdot 
\left\{
\left[\tilde{\mathbf{x}}_{(k)} + \tilde{\mathbf{g}}_{(k)}\right]  - \left[
\tilde{\mathbf{x}}_{(j)}
+
\tilde{\mathbf{g}}_{(j)}\right]\right\}\,,
\end{eqnarray}
where
\begin{equation}
\tilde{\mathbf{g}}_{(k)}\equiv \mathbf{X}_{(k)} \cdot 
\sum_{p=+,\times} \mathbf{e}^p \tilde h_p\,,\quad (k=1,\ldots,N)\,,
\end{equation}
is the GW-induced displacement for detector $k$ in the proper
reference frame (for example of the center of geometry of the entire
configuration).  In this limit, the effect of GW is equivalent to
test-mass displacements, and it is not possible to cancel displacement
noises without completely eliminating the signal. (Canceling the
clock noises alone without eliminating the signal, on the other hand,
{\it is} possible). 
}

From the signals $\tilde\Delta_{(jk)}$, we can construct the following general combination,
\begin{eqnarray}
\label{eq:response}
s(\Omega) &=& {\textstyle \sum_{i \ne j}} D_{(ij)}(\Omega) \tilde \Delta_{(ij)}(\Omega) \nonumber \\
&=& {\textstyle \sum_k \sum_{j \ne k}} A_{(jk)}
\tilde\tau_{(k)} 
+ {\textstyle  \sum_k \sum_{j \ne k} }\mathbf{B}_{(jk)}
\cdot \tilde{\mathbf{x}}_k \nonumber \\
&+& \big[{\textstyle \sum_k \sum_{j \ne k}} \mathbf{G}_{(jk)}e^{-i\Omega \mathbf{e}_z\cdot\mathbf{X}_{(k)}}\big]: {\textstyle \sum_p}\tilde{h}_p \mathbf{e}^p \,,\qquad
\end{eqnarray}
where
\begin{eqnarray}
A_{(jk)} &\equiv& D_{(jk)}e^{i\Omega L_{(jk)}}-D_{(kj)} \,, \\
\mathbf{B}_{(jk)} &\equiv& \left[D_{(jk)}e^{i\Omega L_{(jk)}}+D_{(kj)}\right]\mathbf{n}_{(jk)} \,, \\
\mathbf G_{(jk)} &\equiv& \frac{\left[ A_{(jk)} +\mathbf{e}_z \cdot \mathbf{B}_{(jk)}\right]}
{i\Omega \big[1-[\mathbf{e}_z\cdot \mathbf{n}_{(jk)}]^2\big]}\mathbf{n}_{(jk)}\otimes \mathbf{n}_{(jk)}\,.
\end{eqnarray}
In order to have vanishing noise, we must impose
\begin{eqnarray}
\label{cond1}
\textstyle \sum_{j\ne k} A_{(jk)} =0\,,\qquad \sum_{j\ne k} \mathbf{B}_{(jk)} =0\,,
\end{eqnarray}
which are $N(d+1)$ equations, for the $N(N-1)$ unknowns, $D_{(jk)}$. If $N>d+2$, the number of linear equations is {\it smaller} than the number of unknowns, and we are {\it guaranteed} an $N(N-d-2)$-dimensional null space [of $D_{(jk)}$], in which the noise vanishes. We also need to demonstrate that, the response to GWs, i.e., the third term in Eq.~\eqref{eq:response}, does not vanish identically in this space. 

{\it Relation to interferometry.} Before calculating GW transfer functions for specific configurations, let us argue briefly that our treatment for pulse arrival times can be carried over to laser interferometry. [More detailed investigations will be carried out in Ref.~\cite{CK2}.]  We note: (i) in geometric optics, light rays follow null geodesics,  (ii) optical phase stays constant on phase fronts, and that (iii) interferometry provides a comparison between phases of the local and the arriving remote laser light. From these, we deduce that interference signals correspond to the difference between receipt and emission ``times,'' as indicated by phases of local and remote lasers (which could be noisy), respectively, with an accuracy   limited by quantum fluctuations. 

Specifically, for the light ray from  $(j)$ to $(k)$, we can write:
\begin{equation}
\Phi_{(jk)} = \big[\Phi_{(k)}^{\rm rec.}-\Phi_{(j)}^{\rm emis.} \big]+\delta\Phi^{\rm shot}_{(jk)}\,.
\end{equation}
Here $\Phi_{(k)}^{\rm rec.}$ and $\Phi_{(j)}^{\rm emis.}$ are phases of lasers $k$ and $j$ at light receipt and emission times, respectively --- they can be identified with $\omega_0 t_{(k)}^{\rm rec.}$ and $\omega_0 t_{(j)}^{\rm emis.}$, i.e., proportional to ``time'' indicated by local lasers. This makes our results for time delays apply to the first part of interferometry signal.  The second part, $\delta\Phi^{\rm shot}$, arises from vacuum fluctuations. We call this the shot noise, as in Refs.~\cite{LIGO,LISA}.

 {\it 1-D configurations.}
{Denoting the unit vector along the straight line on which the detectors lie as $\mathbf{n}$, we have
\begin{equation}
\mbox{1-D:}\;  \sum_{j\ne k} \mathbf{G}_{(jk)} = \frac{ \sum_{j\ne k} \left[  A_{(jk)}+\mathbf{e}_z \cdot \mathbf{B}_{(jk)} \right]}
{i\Omega \big[1-[\mathbf{e}_z\cdot \mathbf{n}]^2\big]}\mathbf{n}\otimes \mathbf{n}\,,
\end{equation}
which vanishes whenever Eqs.~\eqref{cond1} are satisfied. Therefore GW signal vanishes whenever all noises are cancelled.  

{\it 2-D, 5-detector configuration.}
The disappointment in 1-D configurations does not
generalize to 2-D configurations. For $d=2$, $N=5$ is the minimum number of detectors. [Note that LISA is formed by 3 spacecraft, and it is natural that LISA cannot provide noise-free timing combinations.] We here study a simple 5-detector network, with 4 of the them lying on the vertices of a square,
and another in the center, as shown in Fig.~\ref{fig1}. Spatial TT coordinates of these 5 detectors, in absence of displacement noises, are given by  
\begin{eqnarray}
\mathbf{X}_{(1)}=(0,0,0)\,, 
\;\mathbf{X}_{(2,3,4,5)}=(\pm L/\sqrt{2},\pm L\sqrt{2},0)\,.
\end{eqnarray}
Apparently there are $20$ timing signals, but since
 $\{3,1,5\}$ ($\{2,1,4\}$) are located on the same
straight line, we 
do not have to include timings between $3$ and $5$ ($2$ and $4$), because
they can be reproduced by combining timings between $3$ and $1$, and  $1$
and $5$ ($2$ and $1$, and $1$ and $4$). We are therefore left with 16
timing signals, which yield a unique noise-free
combination (there are $15$ noises), with:
\begin{eqnarray}
\label{eq:D23}
&&D_{(23)}=D_{(43)}=D_{(45)}=D_{(25)}=z\,,\\
&&D_{(32)}=D_{(34)}=D_{(54)}=D_{(52)}=-z \,, \\
&&D_{(12)}=-D_{(13)}=D_{(14)}=-D_{(15)}=\alpha\,,\\
&&D_{(21)}=-D_{(31)}=D_{(41)}=-D_{(51)}=\beta\,,
\label{eq:D21}
\end{eqnarray}
where
\begin{eqnarray}
\alpha &\equiv & 1 - 1/{\sqrt{2}} + w +{w}/{\sqrt{2}}\,,
\\
\beta &\equiv & -z -{z}/{\sqrt{2}} - wz +{wz}/{\sqrt{2}}
\,,
\end{eqnarray}
and $z \equiv e^{i \Omega L/c}$,$\quad w\equiv e^{i\sqrt{2}\Omega L/c}$. [Factors $z$ and $w$ correspond to time delays by $L/c$ and $\sqrt{2} L/c$, respectively.]

It is easy to note from Fig.~1 that this timing combination gains a minus sign when rotated by $90^\circ$, i.e.,
\begin{equation}
D_{(i'j')}=-D_{(ij)}\,;\; (1,2,3,4,5)'=(1,5,2,3,4)\,.
\end{equation}
We can decompose the noise-free timing combination into a sum of 6 terms, each realizable by conventional equal-arm interferometry and hence has vanishing laser noise; these terms form two groups: 4 of them each realizable by a Sagnac interferometer, and 2 of them each by a Michelson,
 as shown in Fig.~\ref{fig2}. 
 \comment{
 It is straightforward to verify that the 6 interferometers together yield vanishing displacement noise, in the following steps: (i) the motion of $1$ is cancelled within the 4 Sagnac's and the 2 Michelson's, respectively; (ii)  the 4 Sagnac's together only sense motions of 2,3,4 and 5 that are along the diagonals, in the particular mode in which with the same amplitude $2$ and $4$ move away from $1$, while $3$ and $5$ move toward $1$; (iii) this mode of motion will then be cancelled by the Michelson's, when signals from both groups are summed with the appropriate, frequency dependent weight.} (More details regarding this configuration will be provided in Ref.~\cite{CK2}.)

\begin{figure}[t]
\begin{center}
\psfrag{a23}[][]{$+z$}
\psfrag{a32}[][]{$-z$}
\psfrag{a34}[][]{$-z$}
\psfrag{a43}[][]{$+z$}
\psfrag{a45}[][]{$+z$}
\psfrag{a54}[][]{$-z$}
\psfrag{a52}[][]{$-z$}
\psfrag{a25}[][]{$+z$}
\psfrag{a12}[][]{$+\alpha$}
\psfrag{a13}[][]{$-\alpha$}
\psfrag{a14}[][]{$+\alpha$}
\psfrag{a15}[][]{$-\alpha$}
\psfrag{a21}[][]{$+\beta$}
\psfrag{a31}[][]{$-\beta$}
\psfrag{a41}[][]{$+\beta$}
\psfrag{a51}[][]{$-\beta$}
\epsfig{file=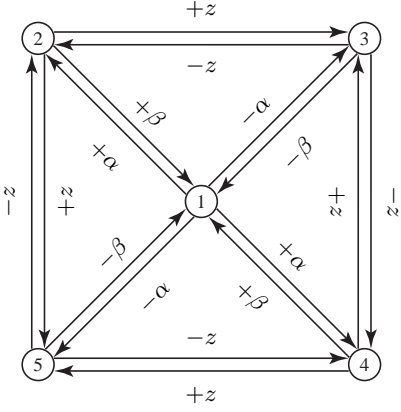, width=0.3\textwidth} 
\end{center}
\caption{\label{fig1} Layout and the noise-free combination for a 5-detector system: one detector at each vertex of a square with side length $\sqrt{2}L$, and another at the center of the square. }
\end{figure}

\begin{figure}[t]
\psfrag{s12}[][]{$+1$}
\psfrag{s13}[][]{$-1$}
\psfrag{s21}[][]{$-zw$}
\psfrag{s31}[][]{$+zw$}
\psfrag{s14}[][]{$+1$}
\psfrag{s15}[][]{$-1$}
\psfrag{s41}[][]{$-zw$}
\psfrag{s51}[][]{$+zw$}
\psfrag{s23}[][]{$+z$}
\psfrag{s32}[][]{$-z$}
\psfrag{s45}[][]{$+z$}
\psfrag{s54}[][]{$-z$}
\psfrag{s25}[][]{$+z$}
\psfrag{s52}[][]{$-z$}
\psfrag{s34}[][]{$-z$}
\psfrag{s43}[][]{$+z$}
\psfrag{m12}[][]{$+\gamma$}
\psfrag{m13}[][]{$-\gamma$}
\psfrag{m21}[][]{$+z\gamma$}
\psfrag{m31}[][]{$-z\gamma$}
\psfrag{m14}[][]{$+\gamma$}
\psfrag{m15}[][]{$-\gamma$}
\psfrag{m41}[][]{$+z\gamma$}
\psfrag{m51}[][]{$-z\gamma$}
\centerline{\epsfig{file=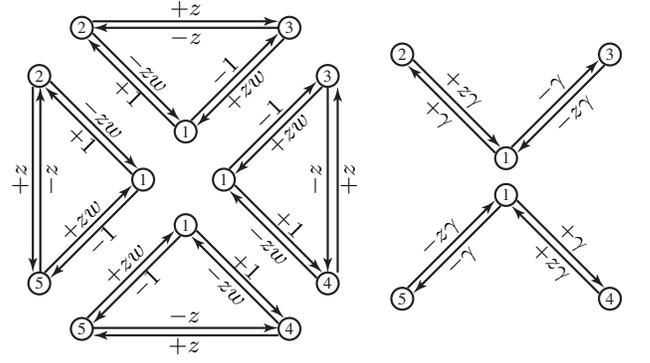, width=0.45\textwidth}}
\caption{\label{fig2} Decomposition of the noise-free timing combination into subsystems realizable by  interferometry: left, four subsystems each realizable by a Sagnac interferometer; right, two subsystems each realizable by a Michelson interferometer.
[We have denoted $\gamma \equiv (1+1/\sqrt{2})(w-1)$.]}
\end{figure}

Writing
\begin{eqnarray}
\mathbf{e}_x &=&
(\cos\theta\cos\phi,\cos\theta\sin\phi,-\sin\theta)\,,\\
\mathbf{e}_y &=& (-\sin\phi,\cos\phi,0)\,, \\
\mathbf{e}_z &=& (\cos\phi\sin\theta,\sin\theta\sin\phi,\cos\theta)\,,
\end{eqnarray}
and $(h_+,h_\times) = (\cos2\alpha,\sin2\alpha)h$, we can write the signal in terms of wave-propagation direction, $(\theta, \phi)$ , wave polarization $\alpha$, angular frequency, $\Omega$, and amplitude $h$,  by inserting Eqs.~\eqref{eq:D23}--\eqref{eq:D21} into Eq.~\eqref{eq:response}. The signal in our noise-free combination is indeed non-vanishing; we obtain analytic results for the transfer function $\mathcal{T}(\Omega) \equiv s/h$, for $\theta=0$,
\begin{eqnarray}
\mathcal{T}_{\theta=0}
=
\frac{i(z-1)}{\Omega}
\!\!\!&[\!\!\!& (2+\sqrt{2})(z-w) \nonumber \\
\!\!\!& +\!\!\!&  (2-\sqrt{2})(zw-1)]\sin(2\phi+2 \alpha)\,,\quad
\end{eqnarray}
and for $\Omega \ll c/L$
\begin{eqnarray}
\label{eq:5det:lf}
\mathcal{T}_{\Omega \ll L/c} \!\!\!& =\!\!\!& \frac{i\Omega^3 L^4\cos\theta}{6c^4}
[2  \cos\theta  \cos2\alpha \sin2\phi  \nonumber \\
&&+  (1+\cos^2\theta)\sin2\alpha\cos2\phi]\,.\;
\end{eqnarray}
In Fig.~\ref{fig3}, we show the root-mean square transfer function, averaged over $\theta$, $\phi$ and the polarization angle $\alpha$:
\begin{equation}
\mathcal{T}_{\rm rms}^2 (\Omega) \equiv   \int_0^{2\pi}  \frac{d\alpha}{2\pi} \int_0^{2\pi}  \frac{d\phi}{2\pi} \int_0^{\pi}  \frac{\sin\theta d\theta}{2} 
 \left|\mathcal{T}(\theta,\phi,\alpha;\Omega)\right|^2 \,.
\end{equation}
In the plot, we verify that $\mathcal{T}_{\rm rms}(\Omega) \sim L/c$ in a frequency band with $\Delta f \sim f$. On the other hand, as we note from Eq.~\eqref{eq:5det:lf} and Fig.~\ref{fig3}, our low-frequency transfer function is $\mathcal{T}\sim \Omega^3$.

{\it Conclusions and Discussions.} In this Letter, we have demonstrated the consistency between displacement-noise cancellation and timing-noise (laser-noise) cancellation in GW detection.  Because of laser-noise cancellation, our schemes do not require laser devices to be fixed with respect to other components of the system, and can {\it in principle} be realized by laser-noise-free interferometry (in our case the combination between four Sagnac and two Michelson interferometers). Such possibility might open a new path toward GW detection.  In particular, the sensitivity of the detector can be made infinitely good, in principle, by increasing laser power, because radiation-pressure noise, which increases with laser power, will be canceled.  [This is more straightforward than using Quantum Non-Demolition schemes, in which the evasion of radiation-pressure noise relies on quantum correlations~\cite{KLMTV}.]

Our example scheme is capable of providing a transfer function,  $\mathcal{T}\sim L/c$ from $h$ to time delay, or $\sim \omega_0  L/c$ from $h$ to optical phase shift, for a frequency band with at least $\Delta f \sim f$, centered around $f \sim c/(2 L)$.  In low frequencies (i.e., $f \ll c/L$), on the other hand, our transfer function goes to $0$ rather steeply, with $\mathcal{T} \sim f^3$. As will be shown in a forthcoming paper~\cite{CK2}, 
this is a fundamental limit for 2-D schemes free from both displacement and timing noises. In 3-D, as we shall see, the best one can do is $\mathcal{T} \sim f^2$, e.g., in a configuration in which detectors occupy vertices of an Octahedron~\cite{CK2}.  Suppose shot noise to dominate, such transfer functions imply noise spectra of $S_h \sim f^{-3}$ in 2-D, and $S_h \sim f^{-2}$ in 3-D.  In comparison, if noisy forces acting on test masses have white spectra at low frequencies (as in LISA), then the uncanceled displacement noise will be $\sqrt{S_h(f)} \sim  f^{-2}$;  in addition, the free-mass Standard Quantum Limit (SQL) scales as $\sqrt{S_h(f)} \sim  f^{-1}$~\cite{SQL}.  
Both spectra grows slower than our shot-noise spectrum when $f \rightarrow 0$, which means rather high laser power (or the use of high-Finesse cavities) will be required, in order to take advantage of  this  displacement-noise-free configuration at low frequencies. 

\begin{figure}[t]
\psfrag{xx}[t][]{$\Omega L/(\pi c)$}
\psfrag{yy}[b][]{$\mathcal{T}_{\rm rms} /(L/c)$}
\centerline{\epsfig{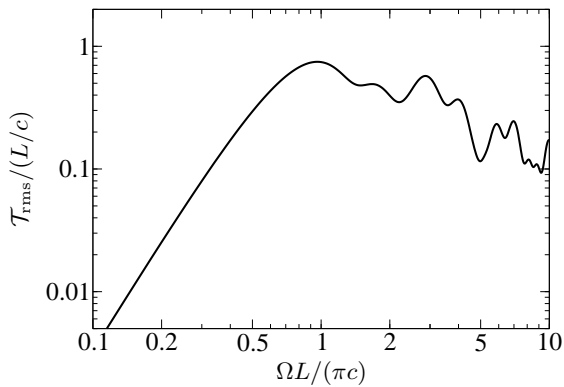}}
\caption{\label{fig3} Root-mean square transfer function of the displacement-noise-free configuration.}
\end{figure}

In generic laser- and displacement-noise-free schemes, we cannot avoid the situation of having detectors which connect to more than two other detectors. [Suppose each detector is connected at most to two other detectors, then the total number of links is no more than $2N$, which is less than the total number of noises, $N(d+1)$, when $d \ge 2$.] For example, in Fig.~\ref{fig1}, each detector on vertices of the square (detectors 2,3,4,5) connects to three other detectors, while the detector at the center (detector 1) connects to all four other detectors. Such schemes cannot be realized solely by using one simple reflective surface for each detector. However, if multiple reflective surfaces were to be combined to realize one idealized detector, relative motions between these surfaces, e.g., due to thermal fluctuations, might not be canceled. The feasibility of implementing displacement-noise-free interferometry is a subject for further research; one interesting possibility is to use different orders of diffractive gratings effectively as mirrors facing different directions, yet sharing the same displacement~\cite{Wise}.

Finally, we note that our analysis is highly idealized. In practice, optical elements must still be vibration isolated, with enough control applied to their translational and rotational degrees of freedom; lasers must also be stabilized in frequency and intensity --- in order to avoid noises arising from non-linear couplings.  In addition, noises generated within each individual optical path, e.g.\ due to scattering, cannot be canceled by our scheme. 

We thank V.B.~Braginsky, C.~Cutler, A.~Pai, E.S.~Phinney, K.~Somiya, K.S.~Thorne and M.~Vallisneri for  useful discussions; in particular, we thank Phinney for pointing out the similarity between our scheme and closure-phase techniques.  Research of Y.C.\ is supported by the Alexander von Humboldt Foundation's Sofja Kovalevskaja Programme and the David and Barbara Groce fund at the San Diego Foundation. Y.C.\ thanks the National Astronomical Observatory of Japan for hospitality and support during his stay.


\begin{thebibliography}{99}
\frenchspacing
\bibitem{KC} S.~Kawamura and Y.~Chen, Phys. Rev. Lett. {\bf 93}, 211103 (2004).
\bibitem{Malik} M.~Rakhmanov, Phys. Rev. D {\bf 71}, 084003 (2005).
\bibitem{LIGO} A. Abramovici et al., Science {\bf 256}, 325 (1992).
\bibitem{LISA}  K.~Danzmann and A.~R\"udiger, Class.~Quantum Grav.~{\bf 20}, S1-S9 (2003).
\bibitem{TDI} M.~Tinto, F.B.~Estabrook and J.W.~Armstrong, Phys.\ Rev.\ D, {\bf 65}, 082003 (2002). 
\bibitem{PC} R.C.~Jennison, Mon. Not. R. Astron. Soc. {\bf 118}, 276 (1958); A.C.S.~Readhead,  and P.N.~Wilkinson, Astrophys. J. {\bf 223}, 25 (1978); T.J.~Cornwell, Science {\bf 245}, 263 (1989).
 \bibitem{CK2} Y.~Chen, S.~Kawamura et al., in preparation.
\bibitem{SQL} See, e.g., V.B. Braginsky and F.Ya. Khalili,  {\it Quantum Measurement,}
edited by K. S. Thorne, Cambridge University Press, Cambridge, England, 1992. 
\bibitem{KLMTV} H.J.~Kimble et al., Phys~.Rev.~{\bf D} 65 022002  (2002). 
\bibitem{Wise} S.~Wise et al., Phys.~Rev.~Lett.~{\bf 95}, 013901 (2005). 
\end{thebibliography}
\end{document}